\documentclass[reprint,superscriptaddress,amsmath,amssymb,aps,twocolumn,prx]{revtex4-2}

\usepackage[T1]{fontenc}
\usepackage{verbatim}
\usepackage{amssymb}
\usepackage{graphicx}
\usepackage{xcolor}
\usepackage{physics}
\usepackage[bookmarks=false,linkcolor=blue,urlcolor=blue,colorlinks,citecolor=blue]{hyperref}
\usepackage{babel}
\usepackage{placeins} 
\usepackage[utf8]{inputenc}

\newcommand{\g}{\ket{\mathrm{gg}}}
\newcommand{\tg}{\mathrm{g}}
\newcommand{\te}{\mathrm{e}}
\newcommand{\0}{\ket{0_L}}
\newcommand{\1}{\ket{1_L}}
\newcommand{\tera}{T_{\mathrm{erasure}}}
\newcommand{\tmeas}{T_{\mathrm{m}}}

\begin{document}

\title{Fast, High-Fidelity Erasure Detection of Dual-Rail Qubits with Symmetrically Coupled Readout}

\author{\mbox{Jimmy Shih-Chun Hung}}
\thanks{These authors contributed equally to this work.}
\affiliation{Amazon Center for Quantum Computing, Pasadena, California 91106, USA}

\author{\mbox{Arbel Haim}}
\thanks{These authors contributed equally to this work.}
\affiliation{Amazon Center for Quantum Computing, Pasadena, California 91106, USA}

\author{\mbox{Mouktik Raha}}
\affiliation{Amazon Center for Quantum Computing, Pasadena, California 91106, USA}

\author{\mbox{Gihwan Kim}}
\affiliation{Amazon Center for Quantum Computing, Pasadena, California 91106, USA}
\affiliation{Institute for Quantum Information and Matter, California Institute of Technology, Pasadena, California 91125, USA}
\affiliation{Thomas J. Watson, Sr., Laboratory of Applied Physics, California Institute of Technology, Pasadena, California 91125, USA}
\author{Ziwen Huang}
\affiliation{Amazon Center for Quantum Computing, Pasadena, California 91106, USA}
\author{\mbox{Ming-Han Chou}}
\affiliation{Amazon Center for Quantum Computing, Pasadena, California 91106, USA}
\author{\mbox{Mitch D'Ewart}}
\affiliation{Amazon Center for Quantum Computing, Pasadena, California 91106, USA}
\author{\mbox{Erik Davis}}
\affiliation{Amazon Center for Quantum Computing, Pasadena, California 91106, USA}
\author{\mbox{Anurag Mishra}}
\affiliation{Amazon Center for Quantum Computing, Pasadena, California 91106, USA}
\author{\mbox{Patricio Arrangoiz Arriola}}
\affiliation{Amazon Center for Quantum Computing, Pasadena, California 91106, USA}
\author{\mbox{Amirhossein Khalajhedayati}}
\affiliation{Amazon Center for Quantum Computing, Pasadena, California 91106, USA}
\author{\mbox{David Hover}}
\affiliation{Amazon Center for Quantum Computing, Pasadena, California 91106, USA}
\author{\mbox{Fernando G.S.L. Brand\~{a}o}}
\affiliation{Amazon Center for Quantum Computing, Pasadena, California 91106, USA}
\affiliation{Institute for Quantum Information and Matter, California Institute of Technology, Pasadena, California 91125, USA}
\author{\mbox{Aashish A. Clerk}}
\affiliation{Amazon Center for Quantum Computing, Pasadena, California 91106, USA}
\affiliation{Pritzker School of Molecular Engineering, University of Chicago, Chicago, Illinois 60637, USA}
\author{\mbox{Alex Retzker}}
\affiliation{Amazon Center for Quantum Computing, Pasadena, California 91106, USA}
\affiliation{Racah Institute of Physics, The Hebrew University of Jerusalem, Jerusalem 91904, Givat Ram, Israel}
\author{\mbox{Harry Levine}}
\affiliation{Amazon Center for Quantum Computing, Pasadena, California 91106, USA}
\affiliation{Department of Physics, University of California, Berkeley, Berkeley, California 94720, USA}
\author{\mbox{Oskar Painter}}
\affiliation{Amazon Center for Quantum Computing, Pasadena, California 91106, USA}
\affiliation{Institute for Quantum Information and Matter, California Institute of Technology, Pasadena, California 91125, USA}
\affiliation{Thomas J. Watson, Sr., Laboratory of Applied Physics, California Institute of Technology, Pasadena, California 91125, USA}

\begin{abstract}
Erasure qubits are a promising platform for implementing hardware-efficient quantum error correction. Realizing the error-correction advantages of this encoding requires frequent mid-circuit erasure checks that are fast, high-fidelity, and scalable. Here, we realize erasure detection with a hardware-efficient circuit consisting of a single readout resonator dispersively and symmetrically coupled to both transmons of a dual-rail qubit. We use this circuit to demonstrate single-shot erasure detection in 384 ns with minimal impact on the dual-rail logical manifold, achieving a residual error per check of $6.0(2) \times 10^{-4}$, with only $8(3) \times 10^{-5}$ induced dephasing per check, and an erasure error per check of $2.54(1)\times 10^{-2}$.
The high degree of matched dispersive readout coupling ($\chi$-matching) within the dual-rail qubit code space also allows us to realize a new modality: time-continuous erasure detection performed in parallel with single-qubit gates. Here we achieve a median $7.2 \times 10^{-5}$ error per gate with $< 1 \times 10^{-5}$ error induced by erasure detection. This demonstrates a reduction in erasure detection overhead as well as a crucial ingredient for soft information quantum error correction. Together, these results establish symmetrically coupled dispersive readout as a fast, hardware-efficient, and scalable component for erasure-based quantum error correction using transmon dual-rail qubits. 
\end{abstract}

\maketitle

\section{Introduction}
Erasure qubits are a promising paradigm for realizing fault-tolerant quantum computing. In this platform, the dominant physical errors are converted into heralded erasure events which can be decoded more easily. Quantum error correction (QEC) protocols with erasure qubits therefore offer higher error thresholds and favorable encoding scalings which translate into significantly reduced overhead~\cite{grassl1997codes,barrett2010fault,wu2022erasure}.
Key to any quantum error correction protocol based on erasure qubits is the design of a qubit encoding and qubit operations that preserve a hierarchy in which erasure errors dominate all other residual forms of error. To this end, recent demonstrations of erasure encoding in neutral atoms~\cite{ma2023high,scholl2023erasure}, superconducting transmons~\cite{levine2024demonstrating, huang2026logical}, cavities~\cite{chou2024superconducting,koottandavida2024erasure, mehta2025bias}, and fluxonium \cite{liu2026converting} have all established high-fidelity erasure qubit operations in which the dominant noise channels are converted to detectable erasure events and a strong error hierarchy is preserved.

An additional key challenge for all the erasure qubit platforms is the need to perform frequent, fast, and high-fidelity mid-circuit erasure checks without disturbing the logical subspace in those cases where an erasure did not occur. In particular, the error induced by the check itself should be mostly convertible to an erasure error, with the induced residual logical error being much smaller and comparable to that of other circuit operations.
Recent studies on superconducting dual-rail qubits have demonstrated high-fidelity mid-circuit erasure checks~\cite{levine2024demonstrating, koottandavida2024erasure, huang2026logical, liu2026converting, degraaf2025mid}. 
These erasure detection schemes, however, come at a cost in terms of hardware overhead, long detection time, or significant errors --- limiting the practicality of the erasure qubit scheme as a whole.

In this work, we demonstrate a technique for erasure detection in transmon dual-rail qubits that is both simple and of high-fidelity, greatly reducing the additional overhead and performance costs of mid-circuit erasure checks. The scheme, proposed in Ref.~\citenum{kubica2023erasure} and schematically displayed in Figs.~\ref{fig:fig1}(a,b), is based on symmetric coupling of a common readout resonator to each transmon qubit in a dual-rail-qubit pair.
Our approach offers three key advantages. First, the symmetrically-coupled readout resonator (SRO) provides hardware-efficient erasure detection in which the resonator serves as both the erasure detector and the logical readout element.
This eliminates the need for an ancillary transmon and its associated control and readout lines from the circuit representing a significant saving for multiplexed architectures. Second, the SRO intrinsically resets on a timescale set by its linewidth $\kappa$ (typically tens to hundreds of nanoseconds) without requiring an additional reset mechanism, thus enabling rapid and repeated erasure checks. Faster resonator reset, if needed, can be implemented using established resonator depletion techniques \cite{mcclure2016rapid, jerger2024dispersive, zhao2025single} to further reduce the inter-check dead time. Lastly, the symmetric coupling enables matched readout coupling within the dual-rail code space ($\chi$-matching) which suppresses measurement-induced dephasing and preserves the coherence of the logical subspace during detection. Using the SRO for single-shot erasure detection, we demonstrate a state-of-the-art total residual error within the logical subspace of $6.0(2) \times 10^{-4}$ per $384$~ns erasure check.
In addition, the high degree of $\chi$-matching also allows one to continuously monitor for erasure events while performing high fidelity dual-rail single-qubit logical operations. Exploiting this, we also perform continuous erasure detection in the strong measurement limit, and achieve a simultaneous single-qubit gate error of $7.2 \times 10^{-5}$, with $<1\times 10^{-5}$ error added by the erasure monitoring. This method not only reduces the time-penalty associated with erasure detection, but also provides a path to more efficient soft information quantum error correction techniques~\cite{raveendran2022soft, pattison2021improved}.

The remainder of the paper is organized as follows. In Sec.~II, we introduce the transmon dual-rail qubit and the symmetrically coupled readout resonator, describing how engineered symmetric coupling achieves the $\chi$-matching condition for erasure detection. In Sec.~III, we characterize the erasure check performance, including extracting residual error per check through interleaved randomized benchmarking. In Sec.~IV, we investigate induced erasure errors during readout. In Sec.~V, we demonstrate continuous parallel erasure detection performed simultaneously with single-qubit gate operations. Finally, in Sec.~VI, we conclude and discuss the outlook for integrating the SRO into erasure-based quantum error correction.

\section{Transmon Dual-Rail and Symmetric Readout}

\begin{figure}[!]
\includegraphics[width = \columnwidth]{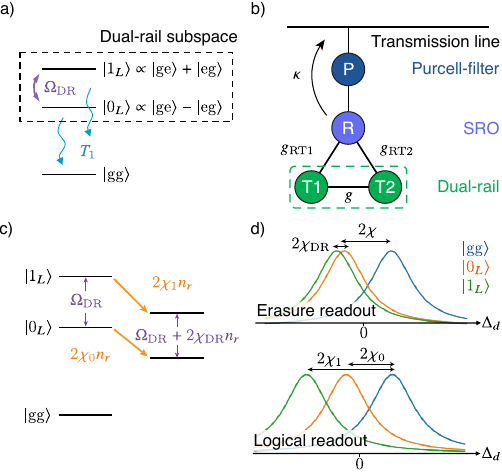}
\caption{\label{fig:fig1}\textbf{Single-shot readout of erasure errors with a symmetrically coupled resonator.}
(a,b) The device consists of two coupled transmons forming the dual-rail qubit in which we encode the qubit within the single excitation subspace. To read out the erasure event, a resonator is symmetrically and dispersively coupled to the dual-rail qubit alongside a single-mode Purcell filter to suppress radiative loss to the readout transmission line.
(c) Dispersive shifts of the resonator to the single excitation states (relative to the $\g$ state) are denoted $2\chi_{\text{0}}$ and $2\chi_{\text{1}}$, with half of the average shift denoted as $\chi = (\chi_{\text{0}}  + \chi_{\text{1}})/2 $. Half the difference between these dispersive shifts is denoted as $\chi_{\text{DR}} = \chi_{\text{1}} - \chi_{\text{0}}$. The modification of the dual-rail energy gap $\Omega_{\text{DR}}$ is given by $2\chi_{\text{DR}}n_r$, where $n_r$ is the average photon number of the resonator.
(d) For ideal erasure detection, the resonator response distinguishes $\0$ and $\1$ from $\g$, but without any difference in response for $\0$ and $\1$. By probing the resonator in this configuration, an erasure check is performed without introducing logical errors. Conversely, by tuning the two transmons off resonance, the same resonator can now be used for three-state readout to additionally differentiate the logical states of the dual-rail.
}\end{figure}

\begin{figure*}[!t]
\includegraphics[width =\textwidth]{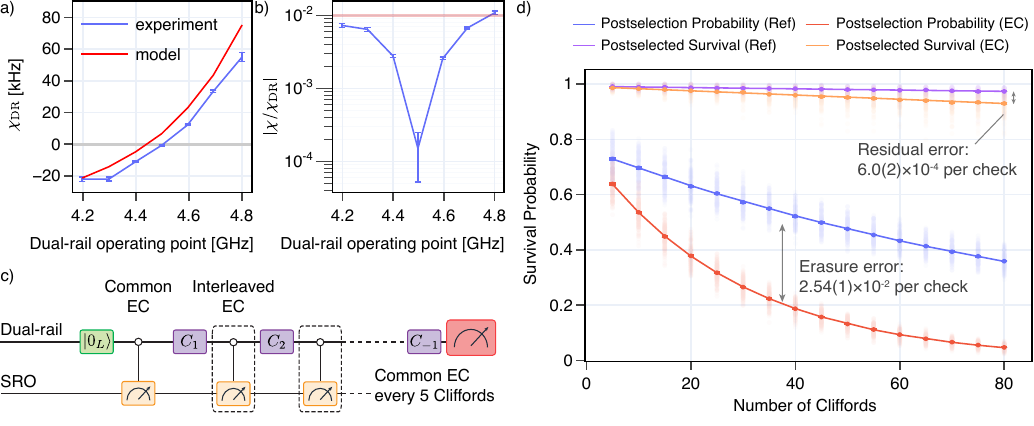}
\caption{\label{fig:fig2}\textbf{$\chi$-matching performance and interleaved randomized benchmarking (ILRB) of erasure checks.}
(a) $\chi_\mathrm{DR}$ is measured as a function of dual-rail operating point (the frequency at which the two constituent transmons are tuned into resonance) over a 600 MHz range. We find the expected $\chi_\mathrm{DR}$ zero-crossing near 4.5 GHz where the logical states are minimally dephased by resonator photon shot noise. The model prediction (red) agrees qualitatively with an offset in the $\chi$-matched frequency likely due to inaccuracies in calibrated circuit parameters.
(b) $|\chi_\mathrm{DR}/\chi|$ quantifies the ability to distinguish the erasure state from logical states while minimally dephasing the logical subspace. A ratio $\lesssim 1 \times 10^{-2}$ (red) is achieved over the entire 600~MHz tuning range characterized. 
(c) Single-qubit ILRB with the erasure check (EC) at the $\chi$-matching point as the interleaved process. Common erasure checks are inserted every 5 Cliffords in both reference and interleaved sequences to postselect against erasure events. Since the erasure check does not occur in parallel with gates, an X gate is inserted during each interleaved check (not shown) for first-order dynamical decoupling of slow frequency noise.
(d) Postselection probability for reference (blue) and interleaved (red) sequences, fitted to exponential decays to 0, yield an erasure error per check of $2.54(1)\times 10^{-2}$. Postselected survival probability with symmetrized readout within the dual-rail subspace for reference sequence (purple) and interleaved sequence (orange) are fitted to exponential decays with an offset of 1/2 to give a residual error per check of $6.0(2) \times 10^{-4}$. Transparent markers show individual random circuit outcomes for each sequence. Including the interleaved check outcomes in postselection further reduces the residual error to $5.4(2) \times 10^{-4}$ at the cost of increasing erasure error to $3.12(1) \times 10^{-2}$ due to the additional false-positive probability.
}\end{figure*}

Our implementation of the dual-rail qubit includes a pair of tunable transmons, coupled and tuned to resonance~\cite{kubica2023erasure} (see Fig.~\ref{fig:fig1}a).
The logical states then correspond to the antisymmetric and symmetric superpositions of the single-excitation states,
\begin{equation}
\begin{split}
    &|1_L\rangle = \frac{1}{\sqrt{2}}\left(|\tg\te\rangle +|\te\tg\rangle\right),
    \\
    &|0_L\rangle = \frac{1}{\sqrt{2}}\left(|\tg\te\rangle -|\te\tg\rangle\right),
\end{split}
\end{equation}
and their frequencies are split by the exchange coupling $\Omega_\mathrm{DR}=2g$ \cite{kubica2023erasure,levine2024demonstrating}. Crucially, in this encoding, the dominant $T_1$ decay channel of the underlying transmons takes the system out of the logical subspace into the joint ground state $\g$. Therefore, if we are able to detect the population in the $\g$ state without disturbing the logical subspace, we are able to convert these amplitude-damping errors of transmons into detectable erasure errors. At the same time, the resonant coupling between the transmons provides passive noise suppression: The dual-rail transition frequency $E_\text{DR} \!\simeq\! \sqrt{(2g)^2 + \delta^2}$ depends only quadratically on the differential frequency fluctuation $\delta$ between the two transmons, strongly suppressing dephasing from low-frequency noise in the underlying transmons frequencies. 
Finally, transitions between logical levels are only induced by the small amount of transmon frequency noise at the gap frequency of $2g \sim 100$ MHz, enabling long logical T1 times.
Overall, the coherence time within the logical subspace can exceed the erasure lifetime by more than an order of magnitude~\cite{kubica2023erasure,levine2024demonstrating}, enabling the use of this system as an \emph{erasure qubit}.

To perform mid-circuit erasure checks in this system we therefore aim at detecting population in the $\g$ level. To this end, we employ a dispersive readout scheme in which both transmons are coupled to a readout resonator~\cite{kubica2023erasure}, as depicted in Fig.~\ref{fig:fig1}(b). Generally, in the dispersive limit, each of the levels $|1_L\rangle$, $|0_L\rangle$, and $\g$ experience a distinct dispersive shift that depends on the readout resonator population, $n_r$. Using the shift of $\g$ as a reference, this is characterized by the relative shifts $2\chi_1$ and $2\chi_0$ for $|1_L\rangle$ and $|0_L\rangle$, respectively. Our goal is for the readout resonator to be able to resolve $\g$ from the logical states, but without having it distinguish between the $|1_L\rangle$ and $|0_L\rangle$, as this would imply measurement-induced dephasing \emph{within} the logical subspace. This would suggest a strategy of maximizing $\chi = (\chi_0 + \chi_1)/2$ while minimizing $\chi_\mathrm{DR} = \chi_1 - \chi_0$.

To make the above statement more quantitative, we can derive the measurement induced dephasing error (i.e. residual error) in the logical subspace for a fixed signal-to-noise-ratio (SNR) in resolving $\g$, accurate for when the system is in its steady state (see Appendix~\ref{sec:ind_deph}):
\begin{align}
    \varepsilon_{\text{deph}} = \frac{\text{SNR}}{12\eta_{\text{eff}}} \left(\frac{\chi_{\text{DR}}}{\chi}\right)^2 \frac{n_{r,\mathrm{DR}}}{n_{r,\mathrm{gg}}},
\label{eq:deph_error}
\end{align}
where $\eta_{\text{eff}}$ is the effective measurement efficiency, and $n_{r,\mathrm{DR}}$ ($n_{r,\mathrm{gg}}$) denotes the steady-state photon number in the resonator during erasure checks when the dual-rail qubit is in its logical (erasure) states. The signal-to-noise ratio is related to the separation error between the readout of logical states and $\g$ through $\varepsilon_\mathrm{sep} = [1 - \mathrm{erf}(\sqrt{\mathrm{SNR}}/2)]/2$.

As anticipated, an optimal readout scheme is one that minimizes the ratio $|\chi_\mathrm{DR}/\chi|$. The induced dephasing error $\varepsilon_\mathrm{deph}$ can be further suppressed by minimizing the ratio $n_{r,\mathrm{DR}}/n_{r,\mathrm{gg}}$ (see Appendix~\ref{sec:ind_deph}),
\begin{equation}
    \frac{n_{r,\mathrm{DR}}}{n_{r,\mathrm{gg}}} = \frac{1 + (2\Delta_\mathrm{d}/\kappa - 2\chi/\kappa)^2}{1 + (2\Delta_\mathrm{d}/\kappa + 2\chi/\kappa)^2},
    \label{eq:selective_darkening}
\end{equation}
where $\Delta_\mathrm{d}$ is the resonator drive detuning. Note that we define the zero of the drive detuning to be centered between the $\g$-state-shifted readout resonance and the average logical subspace shifted resonance (see Fig.~\ref{fig:fig1}d). Driving the readout resonator closer to the shifted resonator frequency associated with the $|\mathrm{gg}\rangle$ state ($\Delta_\mathrm{d} \approx -\chi$) reduces the resonator population build-up when the system is in the logical subspace since the drive is detuned from the resonator by $2|\chi|$ (\emph{selective darkening}), thereby suppressing the induced dephasing. Minimizing Eq.~\eqref{eq:selective_darkening} with respect to $\Delta_\mathrm{d}$ and substituting in Eq.~\eqref{eq:deph_error} results in
\begin{equation}
\begin{split}
\varepsilon_{\text{deph}}^{\text{min}} &= \frac{\text{SNR}}{12\eta_{\text{eff}}} \left(\frac{\chi_{\text{DR}}}{\chi}\right)^2 \left(\sqrt{1 + (2\chi/\kappa)^2} - 2|\chi|/\kappa\right)^2 
\\
&\underset{|\chi|\gg\kappa}{\to}\frac{\text{SNR}}{24\eta_{\text{eff}}} \left(\frac{\chi_{\text{DR}}}{\chi}\right)^2 \left(\frac{\kappa}{2\chi}\right)^2.
\label{eq:deph_error_min}
\end{split}
\end{equation}
As an example, consider a measurement with $ \kappa/2\chi =1$, $\eta_\mathrm{eff}=0.2$, and ${\rm SNR}=19$
(corresponding to a separation error of $\epsilon_{{\rm sep}}\simeq0.1\%$
). 
A $|\chi_{{\rm DR}}/\chi| \sim 1\times 10^{-2}$ leads to a dephasing error per check of  $\epsilon_{{\rm deph}} \sim 4\times 10^{-4}$, a residual error regime relevant for high fidelity fault-tolerant quantum computation \cite{gidney2024magic, jacoby2025magic}.

To achieve a high degree of $\chi$-matching, we design the couplings between the readout resonator and the transmons to be symmetric, $g_\mathrm{RT1}=g_\mathrm{RT2}$ (see Fig.~\ref{fig:fig1}(b)). The readout resonator frequency is detuned by $\Delta$ with respect to the dual-rail operating point, defined by the $|\tg\rangle-|\te\rangle$ transition frequency of the bare transmons. Intuitively, since the readout resonator exerts the same dispersive shift on each of the bare transmons, the logical subspace remains unaffected even upon mixing the states by the coupling $g$. In practice, this intuition holds well in the dispersive limit, but corrections of order $\Delta^{-4}$~\cite{kubica2023erasure} (as well as corrections to the rotating-wave approximation) generally introduce a small but non-zero $\chi_\mathrm{DR}$. These corrections are numerically shown to vanish for a particular value $\Delta=\Delta^\ast$, resulting in a finite operating frequency range around $\Delta^\ast$, where $|\chi_\mathrm{DR}/\chi|$ is sufficiently small.

Figure~\ref{fig:fig2}(a) presents the measured $\chi_{\mathrm{DR}}$ as a function of the dual-rail operating point, from which we identify a zero crossing or "$\chi$-matched" point near 4.5 GHz, where we measured $\chi_{\mathrm{DR}}/2\pi$ to be as small as $-0.7(5)$~kHz.
In Figure~\ref{fig:fig2}(b), we show the modeled and measured ratio $|\chi_\mathrm{DR}/\chi|$, which we find to remain below $\sim1\times 10^{-2}$ over the 600 MHz range characterized.
The ability to maintain a sufficiently low $|\chi_\mathrm{DR}/\chi|$ over a wide tuning range is advantageous as it preserves the tunability of the transmon dual-rail qubit which may assist in avoiding frequency collisions with two-level-system (TLS) defects or spectator modes in a quantum processor \cite{levine2024demonstrating}. See Appendix \ref{sec:chi_dr} for further details of the dispersive shifts characterization.

\section{Benchmarking Erasure Detection}
\label{sec:perf}

\begin{figure}[!]
\includegraphics[width = \columnwidth]{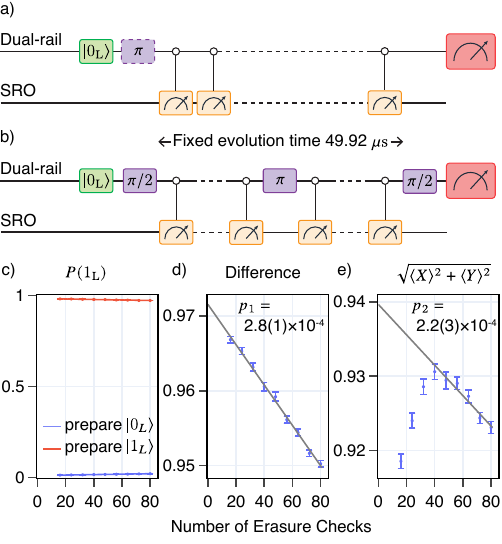}
\caption{\label{fig:fig3}\textbf{Benchmarking of erasure check induced errors.}
(a) We characterize erasure-induced bit-flip by preparing in $ \0/\1$ and inserting a variable number of erasure checks before finally measuring in the Z basis. 
(b) We similarly characterize erasure-induced dephasing by preparing the qubit in a superposition state and performing a Hahn echo sequence with a variable number of checks within a fixed evolution time. 
(c) Postselected probability of measuring $\1$ after preparing in $\0/\1$ and introducing a variable number of erasure checks. 
(d) Postselected bit-flip probability, calculated from the difference of the two curves in (c), is fitted to an exponential decay with an offset of 0 to extract induced bit-flip per check of $p_1 = 2.8(1) \times 10^{-4}$.
(e) Postselected logical phase coherence versus number of erasure checks in a fixed evolution time. Phase coherence initially rises as
more frequent checks improve postselection against relaxation-reheating events that would otherwise dephase the logical qubit.
Induced dephasing per check $p_2 = 2.2(3) \times 10^{-4}$ is extracted by fitting the tail of the decay of phase coherence to an exponential decay to 0. 
}\end{figure}

To evaluate the performance of the erasure detection, we start by calibrating single-shot erasure readout with the dual-rail qubits biased near the $\chi$-matched point at 4.5 GHz. 
The calibrated erasure readout consists of a 384 ns square microwave pulse applied to the SRO, and a 496 ns long integration window which includes 112 ns ($\sim 8.7/\kappa$) of signal during the passive ringdown of the SRO. 
The probe frequency is chosen to match the resonator frequency when the dual-rail qubit is in state $\ket{\mathrm{gg}}$ to minimize readout induced-dephasing. The probe power is then optimized for maximum SNR while limiting measurement induced state transition (MIST)~\cite{sank2016measurement,khezri2023measurement,dumas2024measurement,dai2026characterization} to higher excited states of the dual-rail qubit. See Appendix~\ref{app:ro_cal} for details of readout and classifier calibration.
The calibrated readout achieved an $\text{SNR} = 11.6 (5)$ corresponding to $\varepsilon_{\mathrm{sep}}=0.8\%$.
The false-positive and false-negative probabilities are characterized using a postselected state preparation and measurement experiment \cite{levine2024demonstrating}, where a second measurement ensures high-fidelity state preparation by postselection.
We find that the false-positive and false-negative probabilities are similar to (and likely limited by) the separation error of $\sim 0.8\%$. 

Beyond measurement accuracy, the key performance metrics for an erasure check are the erasure error probability of detectable leakage out of the dual-rail subspace and more importantly the residual error probability of undetected errors within the dual-rail subspace.
For our scheme, we expect the dominant residual errors induced by error checks to be logical dephasing due to the small but non-zero value of $\chi_{\mathrm{DR}}$ (see Appendix \ref{sec:ind_deph}).
To comprehensively characterize errors induced by each erasure check, including idling error contribution, we perform interleaved Clifford randomized benchmarking (ILRB)~\cite{magesan2012efficient} where the interleaved process is an erasure check (see Fig.~\ref{fig:fig2}(c)). Note that here we insert an X gate during the erasure check for dynamical decoupling of slow frequency noise. Such an echoing pulse could otherwise be incorporated into error correction circuits during stabilizer extraction, for instance.

The ILRB protocol is adapted for erasure encoding similar to the method in Ref.~\cite{ma2023high} where we insert periodic mid-circuit erasure checks every 5 Cliffords that are common to both the reference and interleaved sequences to minimize undetected leakage-seepage processes.
In addition to these common erasure checks, the interleaved erasure check provides additional erasure information unique to the interleaved sequence. 
We first postselect on shots where no erasure event is detected by the common checks and measure the postselected survival probability of the random circuits in the dual-rail subspace with symmetrized readout. 
The difference in the decay of postselection probability between reference and interleaved sequences yields an erasure error per check of $2.54(1)\times 10^{-2}$. 
The difference in the decay of postselected survival probability yields a low residual error per check of $6.0(2) \times 10^{-4}$.
Additionally, postselecting on the interleaved check outcomes improves the residual error to $5.4(2) \times 10^{-4}$ at the cost of an increased erasure error of $3.12(1) \times 10^{-2}$ from the extra check's false-positive probability. 
This indicates that rare leakage-seepage processes (e.g. from coherent exchange with spectator modes such as near-resonant qubits or TLS defects) occur on timescales shorter than the common check cadence of 5 Cliffords (every $2.8~\mu\text{s}$) and could be further converted to detectable erasure errors with more frequent mid-circuit checks.

To better understand the error contributions to the measured residual error, we additionally benchmark the error introduced by the erasure check separately from idling errors.
We first characterize the induced bit-flip per check $p_1$ by measuring the loss of postselected polarization with an increasing number of erasure checks within a fixed evolution time of 49.92 $\mu$s when initially preparing the qubit in $\ket{0_L}/\ket{1_L}$ (see Fig. \ref{fig:fig3}(a),(c), and (d)).
By fitting the decay of polarization to an exponential decay  model with an offset of 0, we extract a bit-flip per check of $p_1 = 2.8(1) \times 10^{-4}$.

We then characterize the induced dephasing per check $p_2$ by measuring the loss of postselected phase coherence as we insert an increasing number of erasure checks in a spin-echo measurement with the same fixed evolution time (see Fig.~\ref{fig:fig3}(b)). 
The dual-rail coherence initially improves as we increase the number of erasure checks; this is because these checks and subsequent postselection suppress dephasing from two-step erasure-then-heating processes \cite{levine2024demonstrating}. 
With more checks inserted, the remaining degradation of phase coherence with each additional check reflects only the induced dephasing of the check itself.
We fit this degradation to an exponential decay to 0 to extract an induced dephasing per check of $p_2 = 2.2(3) \times 10^{-4}$ (see Fig.~\ref{fig:fig3}(e)).
The induced dephasing, $p_2 = p_1/2+p_\phi$, includes contributions from both induced bit-flip $p_1$ and induced pure dephasing $p_\phi$, from which we can extract $p_\phi = 8(3) \times 10^{-5}$ (see also Appendix~\ref{app:error_model}).
For our measured level of $\chi_{DR}$, the estimated $p_\phi$ due to photon-shot noise and $\chi$-mismatch is $< 1 \times 10^{-6}$ given the calibrated readout, so our measured $p_\phi$ is likely dominated by other effects which could include near-resonant TLS defects interacting asymmetrically with the dual-rail modes \cite{levine2024demonstrating}.
The small probability of induced bit-flip error may also be a result of rare measurement-induced state transitions \cite{sank2016measurement,khezri2023measurement,dumas2024measurement,dai2026characterization} events to undetected leakage levels incorrectly identified as a bit-flip event. 
We leave a detailed study of bit-flip mechanisms to future work.

\begin{table}[!]
\centering
\begin{tabular}{l c c}
\hline\hline
Source & Expression & Error probability \\
\hline
$T_1$ idling & $\tmeas/3T_1$ & $7.7 \times 10^{-5}$ \\
$T_\phi$ idling & $\tmeas/3T_\phi$ & $1.11 \times 10^{-4}$ \\
Dynamical decoupling (2 X90) & $2\varepsilon_{\text{X90}}$ & $1.8 \times 10^{-4}$ \\
Induced bit-flip & $p_1/3$ & $9.3 \times 10^{-5}$ \\
Induced pure dephasing & $p_\phi/3$ & $2.7 \times 10^{-5}$ \\
\hline
Total accounted & & $4.9 \times 10^{-4}$ \\
Measured residual error & & $6.0(2) \times 10^{-4}$ \\
\hline\hline
\end{tabular}
\caption{Residual error budget per erasure check.}
\label{tab:error_budget}
\end{table}

We estimate that $\sim 80\%$ of the measured residual error of $6.0(2) \times 10^{-4}$ can be accounted for by the error contributions summarized in Table~\ref{tab:error_budget}. Idling error from finite dual-rail coherence is estimated to be $\tmeas/3T_1 = 7.7 \times 10^{-5}$ and $\tmeas/3T_\phi = 1.11 \times 10^{-4}$, single-qubit gate error from the two X90 gates used for dynamical decoupling is estimated to be $2\varepsilon_{\text{X90}} = 1.8 \times 10^{-4}$, and induced errors from the check itself are estimated to be $p_1/3 = 9.3 \times 10^{-5}$ and $p_\phi/3 = 2.7 \times 10^{-5}$. See Appendices~\ref{app:device} and~\ref{app:stab} for device coherence and gate infidelities used in these estimates, and for further details related to the stability of the erasure check performance. 
From this estimated error budget we find that the induced error accounts for less than $20\%$ of the total residual error, with $\chi$-mismatch induced dephasing contributing negligibly, and the idling error during the check dominating the error check performance. 

From these measurements we can also define an erasure noise bias for the erasure check, corresponding to the ratio of erasure error detection probability to residual error probability in the logical dual-rail code space. The demonstrated erasure noise bias of 42(1) for the erasure check process, in combination with the fact that the residual error of the erasure check approaches that of single-qubit gates, highlight the viability of mid-circuit erasure detection using the SRO for erasure-based QEC. Further improvements in erasure check performance can also be anticipated.
Specifically, since the dominant residual error contribution is idling error during the check, improvements in measurement chain quantum efficiency \cite{wang2025high}, optimal control for faster readout \cite{mcclure2016rapid,jerger2024dispersive,zhao2025single}, and better targeting of readout resonator parameters would all yield immediate improvements in erasure measurement time, bringing the residual error of $1 \times 10^{-4}$ within reach.

\section{Readout Induced Erasure Error}

The erasure error probability per check arises from either actual erasure events occurring during the finite readout duration or false-positive detection events due to finite readout accuracy.
The former effect dominates due to our relatively short $\tera$, leading to a $2.54(1)\times 10^{-2}$ erasure error probability per check.
Notably, the $\tera$ while idling ($24.4~\mu\mathrm{s}$ for $\0$ and $21.0~\mu\mathrm{s}$ for $\1$) underpredicts the observed erasure rate. The measured erasure error per check can be up to a factor of two times higher than expected from idle $\tera$ alone, suggesting an accelerated relaxation of the dual-rail excitation in the presence of a readout probe. 

\begin{figure}[!]
\includegraphics[width = \columnwidth]{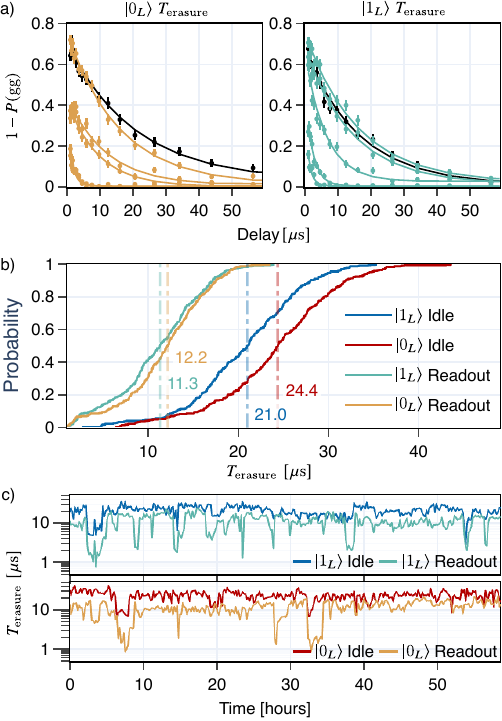}
\caption{\label{fig:fig5}\textbf{Readout-induced degradation of $\tera$.}
 (a) Example $\tera$ decay curves when idling (black) and under a continuous-wave drive at the calibrated probe amplitude (colour) for $\0$ (left) and $\1$ (right). The readout drive can significantly reduce $\tera$, with occasional episodes where substantial decay occurs even at shortest measurement delay times.
 (b) Cumulative distribution of $\tera$ monitored over 48 hours. The median $\tera$ during readout is suppressed by approximately a factor of 2 compared to idling for both logical states. 
 (c) Time series of $\tera$ reveal a general reduction during readout as well as more frequent sporadic dropouts correlated with the appearance of near-resonant TLS defects (see Appendix~\ref{sec:terasure_fluctuation}).
}\end{figure}

To better understand the readout-induced erasure error, we compare the idle $\tera$ to the effective $\tera$ during readout (see Fig.~\ref{fig:fig5}).
The idle $\tera$ is measured with a standard inversion recovery experiment, while the readout $\tera$ is measured with an additional continuous-wave probe on the SRO at the same amplitude and frequency as the calibrated readout.
Over a monitoring period of 2 days, we find the median readout $\tera$ is typically degraded by approximately a factor of two compared to the idle $\tera$, consistent with the measured erasure error per check, up to slow fluctuations of $\tera$ (see Fig.~\ref{fig:fig5}(b, c)). 
While both idle and readout $\tera$ fluctuate over time, the readout $\tera$ exhibits more frequent episodes of significant degradation, which correlate with the presence of near-resonant TLSs independently characterized via spectroscopy (see Appendix~\ref{sec:terasure_fluctuation}).

We attribute the degradation and fluctuation of erasure lifetime under drive to the readout-induced transmon lifetime degradation mechanisms described in Refs.~\cite{sivak2023real, thorbeck2024readout, valles2025optimizing, ziwen2026dispersive}, where Stark shifts and drive-induced broadening enhance the transmon's coupling to frequency-dependent loss channels such as TLS defects.
Notably, our choice of driving at $\Delta_\mathrm{d} = -\chi$ to mitigate $\chi$-mismatched-induced dephasing also minimizes the readout-induced qubit linewidth broadening for a given measurement rate \cite{ziwen2026dispersive}. 
Following these prior studies, we anticipate that control strategies along with improved transmon T1s and fewer strongly-coupled TLSs will enable substantial improvements in induced erasures to reach comparable performance as the high-fidelity QND-ness of state-of-the-art transmon readout.

\begin{figure*}[!]
\includegraphics[width = \textwidth, interpolate=false]{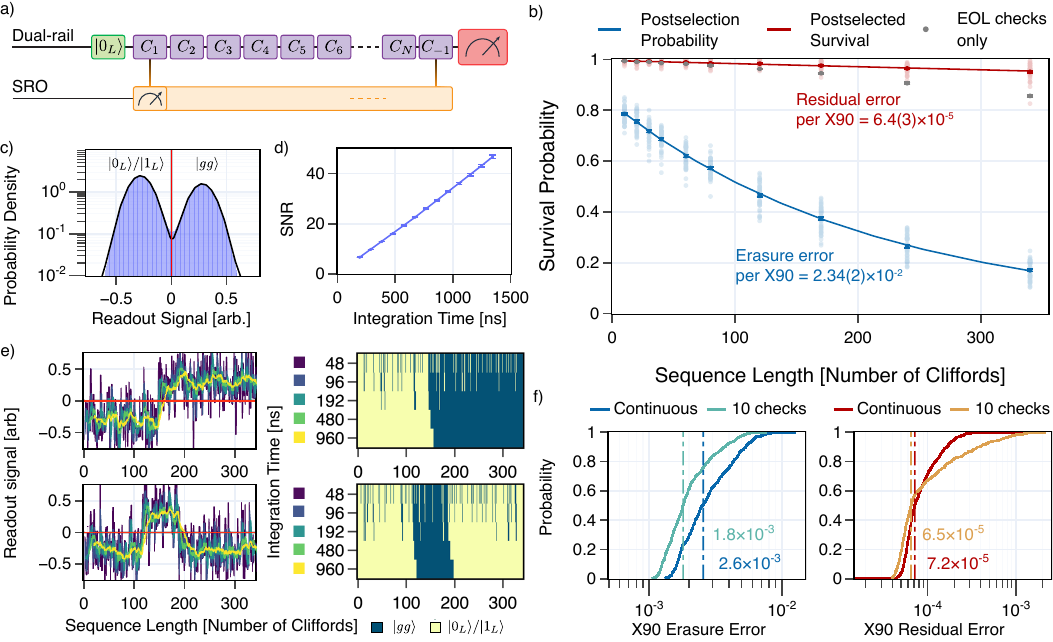}
\caption{\label{fig:fig6}\textbf{Simultaneous operation of dual-rail qubit with continuous erasure checks.}
(a) Single-qubit Clifford fidelity is characterized via randomized benchmarking (RB) while an always-on microwave probe on the SRO continuously detects erasure events.
(b) Postselected survival probability within the dual-rail subspace using only end-of-line (EOL) checks (grey) compared to continuous mid-circuit checks (red). Fitting the continuous case to an exponential decay to 0.5 yields a residual error per X90 of $6.4(3) \times 10^{-5}$ (half the residual error per Clifford).
(c) Histogram of mid-circuit erasure measurements with 480~ns integration time (10 Cliffords), achieving SNR~$= 16.1(1)$ corresponding to false-positive and false-negative rates of $\sim 0.23\%$. The red line indicates the erasure classification threshold.
(d) Linear scaling of measurement SNR with integration time. 
(e) Example continuous erasure detection trajectories at different integration timescales (left-column) and classified outcomes (right-column). (Top-row) an erasure event is detected and the dual-rail remains in $\g$. (Bottom-row) an erasure is followed by a reheating event that returns the dual-rail to the logical manifold, resulting in residual error if uncaught. Erasure check accuracy improves with integration time at the cost of temporal resolution.
(f) Cumulative distribution of X90 gate errors over 5 days consisting of measurements alternating between RB with continuous parallel erasure check (darker traces) and RB with 10 checks spaced evenly throughout the sequence (lighter traces). Parallel erasure checks introduce excess residual error $< 1\times 10^{-5}$ per X90 but reduce the occurrence of sporadic high-error events. This could be due to more effective detection of fast TLS-mediated leakage-seepage processes. Parallel erasure checks also introduce excess erasure error consistent with readout induced degradation of $\tera$.
}\end{figure*}

\section{Continuous Parallel Erasure Detection}
\label{sec:continuous}

While the current 384~ns erasure check demonstrates fast mid-circuit erasure detection, it remains longer than the anticipated dual-rail transmon CZ
gate duration of $\sim$200~ns \cite{kubica2023erasure, huang2026logical}, adding overhead to the QEC cycle time.
Further readout circuit design optimization can bring the erasure check to state-of-the-art transmon readout speeds ($<$100~ns)~\cite{spring2025fast, beaulieu2026fast}, but an alternative approach is to perform erasure checks in parallel with gate operations, eliminating the time overhead entirely.
Such an approach is possible because $\chi$-matching allows the logical subspace to be minimally disturbed during the erasure check.
The continuous nature of the erasure detection can additionally provide fine-grained temporal precision of erasure events, and prior work has shown that improved temporal precision can provide potential benefits in error correction protocols~\cite{gu2025optimizing, chang2025surface}.

We benchmark the performance of continuous parallel erasure detection with a single-qubit Clifford randomized benchmarking (RB) experiment where we probe and measure the SRO response continuously in parallel to detect erasure events (see Fig. \ref{fig:fig6}(a)).
The Clifford gates are each implemented as a pair of X90 gates with virtual Z rotations before, between, and after the gates \cite{mckay2017efficient}. 
The readout probe is chosen to have the same frequency and amplitude as the strong discrete readout benchmarked in Section~\ref{sec:perf}, where hundreds of nanoseconds of integration time provide sufficiently high SNR to distinguish erasure events with high accuracy.
To analyze erasure events with continuous erasure measurement records, we take the simplest approach of integrating the signal over a fixed integration time window and treating each integration window as a distinct single-shot measurement. 
Here we have the flexibility of adjusting the integration window time to achieve a desired measurement SNR (see Fig.~\ref{fig:fig6}(c),(d)). 
We note that this method for detecting erasures could exploit more sophisticated real-time analysis of the continuous measurement record (i.e., more complex filtering) to better extract erasure events on timescales of order $\tera$~\cite{gammelmark2013past, korotkov2001selective, gambetta2007protocols, martinez2020improving}.

For the measurement results presented in Fig.~\ref{fig:fig6}(b), we use a 480 ns long rectangular moving integration window with no overlaps (equivalent to the duration of 10 Cliffords) to detect erasure events, and we postselect on shots where no erasure is detected in any integration window. This integration length achieves an SNR of 16.1(1), equivalent to a separation error of $0.23\%$, which provides sufficiently low false-negative and false-positive probability per check.
Fitting the postselected survival probability of the random Clifford circuit yields a residual error per Clifford of $1.3(1) \times 10^{-4}$. 
Since each Clifford is composed of two X90 gates and error-free virtual Z rotations, the residual error per X90 with concurrent erasure check is half the residual error per Clifford at $6.4(3) \times 10^{-5}$ --- demonstrating high fidelity erasure-detected single-qubit gates.

We can additionally compare to the decay of the survival probability where we only postselect based on the logical readout at the end of the circuit. 
The average survival probability without mid-circuit checks drops significantly, especially for deeper circuits, thus highlighting a key finding: the simultaneous mid-circuit checks enabled by our continuous detection setup are invaluable for larger depths, as they allow one to effectively exclude shots contaminated by leakage and seepage events.  
The left column of Fig.~\ref{fig:fig6}(e) shows examples of continuous measurement records of the readout signal along an optimal quadrature with increasing integration time. Here we see, as expected, that longer integration windows improve the measurement SNR at the cost of a decreased ability to temporally resolve transitions associated with erasure events.
In the right column of Fig.~\ref{fig:fig6}(e) we use a simple threshold for binary classification of erasure events and display the corresponding binary outcome  for different integration times.
Note that the soft information in the left column can be used directly as input to the QEC decoder and is expected to yield superior performance compared to the discrete outcomes shown in the right column, as analyzed, for example, in Ref.~\cite{raveendran2022soft, pattison2021improved}.
The parallel mid-circuit erasure check is able to detect and temporally resolve the location of these events while the circuit is running.
This allows us to effectively exclude shots that would have otherwise contributed to our total residual error.

To make a more comprehensive assessment of single-qubit (1Q) gate performance in the presence of simultaneous parallel erasure monitoring, we measured 1Q gate errors over 5 days, interleaving experiments using two different approaches to erasure postselection. In one case we used continuous parallel erasure detection, and in the other we used a fixed number of discrete erasure checks interleaved with the Clifford gates.
In the case with discrete erasure checks, we determine the postselected gate error using a set of 10 discrete mid-circuit erasure checks spaced evenly throughout the circuit. In this scenario, the erasure check infidelity does not contribute to the extracted gate error~\cite{levine2024demonstrating}, and the discrete checks are sufficiently frequent such that there is at most a $\tera / 18$ gap between each check for the longest sequences.
Fig.~\ref{fig:fig6}(f) shows the cumulative distributions of X90 gate errors for the two configurations: continuous parallel erasure checks (darker traces) and discrete mid-circuit checks (lighter traces).
The erasure error per gate is elevated from $1.8 \times 10^{-3}$ to $2.6 \times 10^{-3}$ under parallel erasure check, consistent with our previous observation of $\tera$ degradation during readout.
The residual error distributions with and without parallel erasure checks have similar median values of $7.2 \times 10^{-5}$ and $6.5 \times 10^{-5}$, respectively.
This slight degradation is consistent with the expected contribution from the small induced dephasing and bit-flip rates characterized in  Section~\ref{sec:perf}, which amount to $(p_1/3 + p_\phi/3) \sim 7 \times 10^{-6}$ when scaled to the gate duration of 24~ns.

The cumulative distributions additionally reveal a striking feature. The baseline residual gate error distribution without continuous erasure detection exhibits a long tail of sporadically high errors, which we attribute to fluctuating near-resonant TLS defects present in this device.
In stark contrast, this tail is notably absent from the distribution measured with parallel continuous erasure checks. 
This suggests that occasionally faster leakage-seepage mechanisms (such as coherent exchange with fluctuating defect modes) can produce residual errors that are effectively suppressed by more frequent erasure detection, as is the case with continuous checks. This is consistent with our observation from interleaved RB characterization in Section~\ref{sec:perf}, where the use of more frequent erasure detection information from the interleaved checks can similarly lead to further erasure conversion of residual error.

\section{Conclusion and Outlook}
We demonstrated in this work high-fidelity mid-circuit erasure detection of a transmon dual-rail qubit using a symmetrically coupled readout resonator. The engineered symmetric coupling achieves $\chi$-matching with $|\chi_{\text{DR}}/\chi| < 1\times 10^{-2}$ over a 600 MHz bandwidth, enabling single-shot erasure detection in 384 ns with a separation error of 0.8\% and residual error per check of $6.0(2) \times 10^{-4}$. 
The $\chi$-matched readout further enabled us to demonstrate continuous erasure detection executed in parallel with 1Q logical operations, achieving an X90 gate infidelity of $7.2 \times 10^{-5}$, of which only $7 \times 10^{-6}$ error probability is attributable to the detection itself. 

Continuous parallel erasure detection introduces new opportunities for improving QEC performance. First, the parallelization of erasure detection reduces the circuit overhead. Second, continuous erasure monitoring provides soft information and fine-grained temporal resolution of the erasure events, offering key advantages over discrete measurements. In particular, it provides the probabilities of erasure rather than binary outcomes, which can be efficiently exploited by the decorder \cite{raveendran2022soft,pattison2021improved}. Furthermore, the inherent temporal resolution mitigates error propagation by enabling more precise localization of erasure events, which has been shown to improve QEC performance \cite{gu2025optimizing, chang2025surface}. This last point was demonstrated in Ref. \cite{gu2025optimizing} that explored the advantage of increasing the frequency of erasure checks, of which continuous erasure detection is the limiting case. 

We have also analyzed the different contributions to the measured residual and erasure errors, which point to clear paths towards further improvement in erasure check performance.
The residual error per check is dominated by echo pulse errors and idling errors during the measurement window, with measurement-induced contributions accounting for less than 20\% of the total.
In a realistic error correction circuit with cycle time comparable to or longer than the erasure check, dynamical decoupling can be incorporated into the stabilizer circuit, and additional decoupling pulses would not be required specifically for the erasure checks.
In terms of idling errors, optimization of the system for faster readout through higher quantum efficiency amplification~\cite{wang2025high} and optimal control \cite{mcclure2016rapid, jerger2024dispersive, zhao2025single, wang2025high} represents a promising path towards further reduction in residual errors.
Additionally, induced dephasing from residual $\chi$-mismatch, while already negligible, can be further suppressed by increasing $\chi/\kappa$ to better leverage selective darkening. 
The induced bit-flip contribution, potentially from rare measurement-induced state transitions~\cite{sank2016measurement, khezri2023measurement, dumas2024measurement, dai2026characterization}, may be reduced by operating at lower photon numbers, by extending the classifier to detect higher excited states for leakage reduction (see Appendix~\ref{app:leak}), or by designing for operation in a highly-detuned regime of dispersive readout \cite{kurilovich2025high, dixit2026millimeter}.
Finally, the broad effort towards improving transmon $T_1$ times, including the reduction of TLS density through materials and fabrication advances \cite{bland2025millisecond}, translates naturally into the dual-rail architecture to directly reduce the erasure error per check.

In addition to its performance merits, the SRO circuit also eliminates the need for dedicated ancilla transmons in erasure detection, thus representing a hardware savings that becomes increasingly significant at the scale of a full error-correcting code. 
In this approach, the same resonator employed for erasure detection serves multiple roles.
While the $\chi$-matched condition deliberately hides logical state information from the resonator, this degeneracy can be broken on demand either by introducing a controlled $\chi$-mismatch through dynamic detuning of one transmon via its flux-tunable frequency (cf. Fig.~\ref{fig:fig1}d), or by mapping logical states onto different excitation numbers (\emph{e.g.,} $\1 \rightarrow \ket{11}$) before readout, enabling logical state discrimination via the SRO resonator.
The resonator also naturally distinguishes the $n = 0$ (erasure), $n = 1$ (logical), and $n \geq 2$ (higher-excitation leakage such as $\ket{\text{ee}}, \ket{\text{gf}},...$) excitation-number sectors, enabling detection of leakage beyond the single-excitation subspace (see Appendix~\ref{app:leak}). 
The SRO thus serves as a compact multi-purpose element for erasure detection, logical readout, and leakage detection, all without additional circuitry.

\begin{acknowledgments}
We thank Arne Grimsmo and Akshay Koottandavida for helpful discussions. We also thank the staff from across the Amazon Center for Quantum Computing that enabled this project.
\end{acknowledgments}

\FloatBarrier
\appendix 
\section{Device}
\label{app:device}
\setlength{\tabcolsep}{12pt} 
\begin{table}
    \centering
    \begin{tabular}{ccc}
        \hline
        \hline
        Parameter & Value & 
        \begin{tabular}{@{}c@{}} Number of \\ Samples \end{tabular} \\ 
        \hline
        $T_1$ & 2176 $\mu \text{s}$ &  779 \\ 
        CPMG-16 $T_2$ & 1116 $\mu \text{s}$ & 810\\
        CPMG-1 $T_2$ & 471 $\mu \text{s}$ &  809 \\
        $\tera^{\ket{0_L}}$ & 27 $\mu \text{s}$ &  779\\ 
        $\tera^{\ket{1_L}}$ & 24 $\mu \text{s}$ & 779\\ 
        X90 residual error & $9\times 10^{-5}$ & 5526 \\
        $\eta_{\text{eff}}/\eta_{\text{ideal}}$ & 25(1) \% & \\
        $\kappa_r^{\g}/2\pi$ & 12.4(2) MHz& \\
        $\kappa_r^{\1}/2\pi$ & 10.5(1) MHz& \\
        $\chi/2\pi$ & -4.25(3) MHz&\\ 
        $\chi_{\text{DR}}/2\pi$ & -0.7(5) kHz& \\ 
        $\Omega_{\text{DR}}/2\pi$ & 94.150(2) MHz& \\
        $f_r^{|\mathrm{gg}\rangle}/2\pi$ & 6.7916(6) GHz&  \\
        \hline
        \hline
    \end{tabular}
    \caption{Summary of device parameters at $\chi$-matched operating point (4.5 GHz). Coherence and fidelity properties are reported as median of results over 5 days without recalibration of the system.}
    \label{tab:device_params}
\end{table}

The experimental device consists of a superconducting circuit implementing the components shown in Fig.~\ref{fig:fig1}(a) where two tunable transmons spanning a tuning range of $\sim$2.5-5.5 GHz form the dual-rail qubit. The pair of transmons are symmetrically and strongly coupled to the readout resonator with a $\chi$-matched point designed and characterized near 4.5 GHz. See Table~\ref{tab:device_params} for device parameters and coherence at the $\chi$-matched point. Purcell loss through the resonator is mitigated by a single-mode Purcell filter. We note that the difference in resonator effective linewidth for $|\mathrm{gg}\rangle$ and $|0_L\rangle/|1_L\rangle$ is a result of the parameter regime of the readout-Purcell system where the readout-Purcell coupling approaches the Purcell resonator external coupling \cite{swiadek2024enhancing}. Finally, the measurement quantum efficiency $\eta_{\text{eff}}$ is characterized with the method outlined in Ref.~\cite{bultink2018general} adapted to the measurement SNR and induced dephasing between $|\mathrm{gg}\rangle$ and $|1_L\rangle$. As we are using a phase-preserving amplification chain with the TWPA as the first stage amplifier, quantum limited efficiency is $\eta_{\text{ideal}} = 50\%$. 

\section{Details on Induced Dephasing and Selective Darkening}
\label{sec:ind_deph}

In this appendix, we derive the measurement-induced dephasing error within the logical subspace due to the dispersive shift mismatch $\chi_{\text{DR}}$, and show how the drive detuning can be optimized to minimize this error.

The effective low-energy Hamiltonian for the dual-rail - SRO system (see Fig.~\ref{fig:fig1}(a)) is given by
\begin{align}
H &= \Big[(\chi\! -\! \chi_{\text{DR}})\ket{0_L}\!\bra{0_L} +\! (\chi\! +\! \chi_{\text{DR}})\ket{1_L}\!\bra{1_L} \nonumber \\
&\quad -\! \chi\ket{\text{gg}}\!\bra{\text{gg}}\Big] a^\dagger a + \varepsilon_\mathrm{d} (a^\dagger\! +\! a) + \Delta_\mathrm{d} a^\dagger a,
\label{eq:hamiltonian}
\end{align}
where $\chi = (\chi_0 + \chi_1)/2$ is the average dispersive shift of the two logical states with respect to the ground state, $\chi_{\text{DR}} = \chi_1 - \chi_0$ is their difference, $\varepsilon_\mathrm{d}$ is the drive amplitude, and $\Delta_\mathrm{d}$ is the drive detuning from the center frequency.

Depending on the state of the dual-rail system ($\0$, $\1$, or $\ket{\text{gg}}$), the resonator evolves to one of three steady states:
\begin{align}
\alpha_0^s &= \frac{-i\varepsilon_\mathrm{d}}{\kappa/2 + i(\Delta_\mathrm{d} + \chi - \chi_{\text{DR}})}, \label{eq:alpha0} \\
\alpha_1^s &= \frac{-i\varepsilon_\mathrm{d}}{\kappa/2 + i(\Delta_\mathrm{d} + \chi + \chi_{\text{DR}})}, \label{eq:alpha1} \\
\alpha_{\text{gg}}^s &= \frac{-i\varepsilon_\mathrm{d}}{\kappa/2 + i(\Delta_\mathrm{d} - \chi)}, \label{eq:alphagg}
\end{align} where $\kappa$ is the resonator decay rate. Because $\chi_{\mathrm{DR}}$ is much smaller than both $\chi$ and $\kappa$, the difference between $\alpha_0^s$ and $\alpha_1^s$ is much smaller than their separation from $\alpha^s_{\text{gg}}$. The steady-state resonator photon numbers for the logical and erasure subspaces are, respectively,
\begin{align}
    n_{r,\mathrm{DR}} \approx \frac{\varepsilon_\mathrm{d}^2}{[(\kappa/2)^2 + (\Delta_\mathrm{d} +\chi)^2]},\nonumber\\
    n_{r,\mathrm{gg}} = \frac{\varepsilon_\mathrm{d}^2}{[(\kappa/2)^2 + (\Delta_\mathrm{d} -\chi)^2]}.
\end{align}

The measurement rate at steady state, which characterizes the rate of information gained about whether the system is in the logical subspace or the erasure state, is given by \cite{sete2015quantum}
\begin{align}
\Gamma_{\text{meas}} &= \frac{\text{SNR}}{T} = 2\kappa\eta_{\text{eff}} |\delta\alpha|^2 \nonumber \\
&= \frac{2\kappa\eta_{\text{eff}} |\varepsilon_\mathrm{d}|^2 (2\chi)^2}{[(\kappa/2)^2 + (\Delta_\mathrm{d} + \chi)^2][(\kappa/2)^2 + (\Delta_\mathrm{d} - \chi)^2]},
\label{eq:meas_rate}
\end{align}
where $\eta_{\text{eff}}$ is the effective measurement efficiency and $|\delta\alpha| = |\alpha_{\text{gg}}^s - (\alpha_0^s + \alpha_1^s)/2|$ is the separation between the erasure and logical steady states in phase space.

The dephasing rate within the logical subspace, arising from the different resonator steady states for $\0$ and $\1$, is given by~\cite{gambetta2026qubit}
\begin{equation}
\begin{split}
\Gamma_{\text{deph}} &= -2\chi_{\text{DR}} \text{Im}[\alpha_0^s (\alpha_1^s)^*] =2|\varepsilon_\mathrm{d}|^2 \chi_{\text{DR}}^2 \kappa \\
&\times \frac{1}{[(\kappa/2)^2 + (\Delta_\mathrm{d} + \chi + \chi_{\text{DR}})^2]}
\\
&\times\frac{1}{[(\kappa/2)^2 + (\Delta_\mathrm{d} + \chi - \chi_{\text{DR}})^2]}.
\label{eq:deph_rate}
\end{split}
\end{equation}
For QND erasure detection, the regime of interest is $\Delta_\mathrm{d} \leq 0$ and $|\chi_{\text{DR}}| \ll |\chi|$, for which we can approximate
\begin{equation}
\Gamma_{\text{deph}} \simeq \frac{2|\varepsilon_\mathrm{d}|^2 \chi_{\text{DR}}^2 \kappa}{[(\kappa/2)^2 + (\Delta_\mathrm{d} + \chi)^2]^2}.
\label{eq:deph_approx}
\end{equation}

This yields an important figure of merit: the ratio of the induced dephasing rate to the measurement rate,
\begin{align}
\frac{\Gamma_{\text{deph}}}{\Gamma_{\text{meas}}} &= \frac{1}{4\eta_{\text{eff}}} \left(\frac{\chi_{\text{DR}}}{\chi}\right)^2\frac{n_\mathrm{r,DR}}{n_\mathrm{r,gg}}.
\label{eq:ratio}
\end{align}
Alternatively, we can express the induced dephasing error for a fixed SNR, yielding Eq.~\eqref{eq:deph_error} of the main text:
\begin{equation}
\varepsilon_{\text{deph}}=\frac{\Gamma_\mathrm{deph}T}{3} = \frac{\text{SNR}}{12\eta_{\text{eff}}} \left(\frac{\chi_{\text{DR}}}{\chi}\right)^2 \frac{n_{r,\mathrm{DR}}}{n_{r,\mathrm{gg}}},
\label{eq:deph_error_full}
\end{equation}
where $n_{r,\mathrm{DR}}/n_{r,\mathrm{gg}}$ is a function of $\Delta_\mathrm{d}$, $\chi$, and $\kappa$.

Finally, minimizing $\varepsilon_{\text{deph}}$ with respect to the drive detuning $\Delta_\mathrm{d}$ results in Eq.~\eqref{eq:deph_error_min} of the main text, where the optimal drive detuning is $\Delta_\mathrm{d}^{\text{min}} = \text{sgn}(\chi) \sqrt{(\kappa/2)^2 + \chi^2}$. Examining the different regimes of $2|\chi|/\kappa$,
\begin{equation}
\varepsilon_{\text{deph}}^{\text{min}} = \frac{\text{SNR}}{12\eta_{\text{eff}}} \left(\frac{\chi_{\text{DR}}}{\chi}\right)^2 \times
\begin{cases}
1, & 2|\chi|/\kappa \ll 1 \\[6pt]
(\sqrt{2} - 1)^2 \approx 0.17, & 2|\chi|/\kappa = 1 \\[6pt]
\dfrac{1}{2}\left(\dfrac{\kappa}{2\chi}\right)^2, & 2|\chi|/\kappa \gg 1.
\end{cases}
\label{eq:regimes}
\end{equation}
demonstrates that \emph{selective darkening}, achieved by operating in the resolved regime $|\chi| \gg \kappa/2$ and tuning the drive to $\Delta_\mathrm{d} \to -\chi$, suppresses the dephasing error by an additional factor of $(\kappa/2\chi)^2$. In this regime, the resonator remains nearly empty for the logical states while becoming populated for the erasure state, thereby minimizing the measurement back-action on the logical subspace while maintaining strong discrimination of erasure events.

\section{Dispersive Shift Characterization}
\label{sec:chi_dr}
\begin{figure}[!]
\includegraphics[width = \columnwidth]{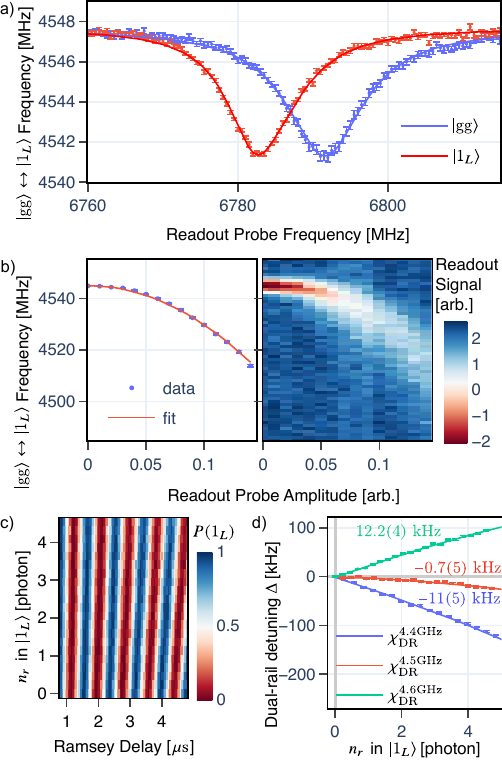}
\caption{\label{fig:fig_a1}\textbf{Dispersive shift and photon number calibration.}
(a) Spectroscopy of $|\mathrm{gg}\rangle\leftrightarrow|1_L\rangle$ transition starting in either $|\mathrm{gg}\rangle$ or $|1_L\rangle$ with variable-frequency readout drives allows us to extract the resonator effective linewidth and $\chi_1$ simultaneously.
(b) Spectroscopy of $|\mathrm{gg}\rangle\leftrightarrow|1_L\rangle$ transition starting in $|1_L\rangle$ with variable drive amplitude to more exactly connect room temperature signal amplitude $A_{ro}$ to average photon number when the qubit is in logical subspace $n_{r,\mathrm{DR}}$. 
(c) $\chi_{\mathrm{DR}}$ is characterized by measuring the dual-rail energy gap detuning as a function of steady state photon occupation in the resonator using Ramsey interferometry.
The example shown for the dual-rail is at 4.6 GHz with $\chi_{\mathrm{DR}}/2\pi = 12.2(4)$ kHz. 
(d) We extract and report the slope of detuning to $n_{r}$ at low photon number for $\chi_{\mathrm{DR}}$ as higher order nonlinear dispersive shifts starts to take effect at photon number higher than those used for readout. For example, we measure $\chi_{\mathrm{DR}}/2\pi = -0.7(5)$ kHz at a near $\chi$-matched operating point of 4.5 GHz.
}\end{figure}

Characterization of the $\chi$-mismatch $\chi_{\mathrm{DR}}$ requires the calibration of the average photon number $n_{r}$ of the SRO when driven into steady state of the readout probe. 
We first characterize the dispersive shift of $|1_L\rangle$ relative to $|gg\rangle$ ($\chi_1$) and photon number calibration using standard transmon methods relying on the ac Stark shift of the $|gg\rangle\leftrightarrow|1_L\rangle$ transition  $\Delta_1 = 2\chi_1n_{r} = 2\chi_1(CA_{ro})^2$ where $A_{ro}$  is the probe DAC amplitude and $C$ is  the scaling factor connecting signal power at room temperature to $n_{r}$. 
Following the method presented in \cite{sank2025system}, we measure the ac Stark shift with the qubit prepared in either $|gg\rangle$ or $|1_L\rangle$ while varying the readout probe frequency at a fixed probe amplitude. 
Fitting the Stark shifted transition frequency to a pair of Lorentzians, we extract both the resonator linewidths and dispersive shift $\chi_1$ (see Fig.~\ref{fig:fig_a1}(a). 
With the readout drive frequency resonant to the dispersively shifted resonator frequency $f_r^{|1_L\rangle} = f_r^{|\text{gg}\rangle} + 2\chi_1$, we additionally initialize in $\1$ and sweep the drive amplitude to more carefully calibrate the quadratic dependence of $n_{r,\mathrm{DR}}$ to $A_{ro}$ (see Fig.~\ref{fig:fig_a1}(b)). 

With the photon number calibration, we can then measure the dual-rail logical states detuning with a Ramsey sequence between the logical states (postselected on remaining in logical subspace) under a variable strength readout probe at $f_r^{|1_L\rangle}$ to extract $\chi_{\mathrm{DR}}$ (see Fig.~\ref{fig:fig_a1}(c), (d)).
The dispersive shift $\chi_{\mathrm{DR}}$ can be nonlinear with respect to $n_{r,\mathrm{DR}}$ \cite{kubica2023erasure} which is more apparent near $\chi$-matched point.
As such, we fit $\Delta_{\mathrm{DR}}(n_{r,\mathrm{DR}})$ to a degree-2 polynomial $2\chi_{\mathrm{DR}}n_{r,\mathrm{DR}}+2\chi_{\mathrm{DR}}'n_{r,\mathrm{DR}}^2$ and report the linear coefficient $\chi_{DR}$ which is most relevant to the small-number-of-photon regime of our calibrated readout where $n_{r,\mathrm{DR}} \sim 1$. 
At the near $\chi$-matched point at 4.5 GHz, we find  $\chi_{\mathrm{DR}}/2\pi = -0.7(5)$ kHz and a similar small higher order dispersive shift of $\chi'_{\mathrm{DR}}/2\pi = -0.7(2)$ kHz. 

\section{Readout Calibration and Classification}
\label{app:ro_cal}
Erasure readout is calibrated to maximize SNR while minimizing measurement duration and observed MIST.  The optimized probe pulse is 384 ns boxcar waveform with a 496 ns long matched filter integration kernel \cite{ryan2015tomography} which allows the resonator plenty of time ($\sim5.5/\kappa$) to deplete naturally. We note that the measurement time can be further improved with a multi-step probe pulse \cite{bultink2018general,mcclure2016rapid, zhao2025single} or more complex pulse parametrization \cite{jerger2024dispersive}. The readout frequency is chosen to be resonant to $f_r^{|\mathrm{gg}\rangle}$ to minimize induced dephasing as described in Appendix~\ref{sec:ind_deph}. This choice of drive frequency also minimizes readout induced erasure due to coupling to off-resonant loss channels such as a TLS since the readout drive asymmetrically broadens the $|0_L\rangle/|1_L\rangle\rightarrow|\mathrm{gg}\rangle$ transitions when driven resonant to $f_r^{|0_L\rangle/|1_L\rangle}$ compared to $f_r^{|\mathrm{gg}\rangle}$ \cite{ziwen2026dispersive}. Finally the probe amplitude is maximized without inducing measurable probability of MIST events. Since the SRO implements close to an excitation number resolving measurement, the same readout resonator is used to provide visibility to MIST events similar to the readout of higher transmon levels with dispersively coupled resonator \cite{wang2025probing}. The erasure check is repeated up to 1000 times back-to-back to amplify the MIST error probability and we maximize the pulse amplitude while making sure the readout signal lies within the distributions corresponding to erasure and logical subspace only. 

To classify the erasure readout, we use a simple threshold at the midpoint between the mean of the two readout signal distributions which leads to unbiased false-positive and false-negative probabilities limited by the separation error from mixing of the Gaussian readout signal distribution due to finite SNR. While not explored in this experiment, direct dispersive readout of erasure events with the SRO offer an additional degree of flexibility compared to ancilla based erasure check as the balance of false-positive and false-negative can be tuned and traded off with a simple adjustment of classification threshold for the application at hand. In comparison, the false-negative probability of an ancilla based erasure check is typically limited by energy relaxation events of the ancilla qubit as one would preferentially conditionally excite the ancilla when the dual-rail is erased to minimize induced dephasing while the qubit remains in logical subspace. Techniques to use two-photon excitation of ancilla \cite{jerger2024dispersive} can mitigate the false-negative due to ancilla relaxation but the ancilla relaxation induced false-negative is nonetheless a dominant and non-tunable error compared to SRO based direct dispersive readout. 

\section{Details of ILRB}
To evaluate the error of the erasure check operation in the logical subspace, we benchmark the erasure check (including the idling error during the finite erasure check duration) with ILRB experiments. 
The standard ILRB protocol is complemented in the same way as Ref.~\cite{ma2023high} with periodic mid-circuit erasure checks (chosen to be sufficiently frequent at every 5 Cliffords) to postselect away shots where an erasure event occurred. 
In this way, we are comparing the survival probabilities of the interleaved and reference sequences conditioned on the dual-rail remaining in the logical subspace. 
The frequent checks allow leakage events to be detected almost immediately with low false-negative error of $0.89(3)\%$. The interleaved circuit compared to reference therefore has a missed erasure fraction of $5\times 2.54\% \times 0.89\% = 0.11 \%$ between each common erasure check. 
As such, we expect minimal impact on postselected survival probability of ILRB sequences due to leakage-seepage events especially when additionally considering the low probability of seepage following missed erasures due to the significantly lower heating rate of the erasure-logical transitions at mK qubit temperature similar to the condition of Ref.~\cite{mehta2025bias}. Specifically, we measure the equilibrium thermal excitations of the individual transmons $p_{\text{equil}}$ to be $0.23(4)\%$ which corresponds to a slow heating timescale of $T_{\text{heat}} \approx \tera/p_{\text{equil}} =$ 13 ms.

As noted in Section \ref{sec:perf}, we can additionally use the erasure check information of the interleaved checks that is only available in the interleaved circuit to additionally postselect against erasure events that may not be caught by the sparse erasure check common to both sequences. 
Comparison of the residual error extracted with and without the use of additional and frequent erasure check information from the interleaved sequence shows a small but measurable difference in error rate of $0.060(2)- 0.054(2) = 0.006(4) \%$. 
This suggests that despite the low missed erasure fraction and slow heating timescale, there are still potentially very rare  seepage processes faster than the heating timescale possibly from faster coherent exchanges with TLS defects.

\section{Induced Error Model and Residual Gate Infidelity}
\label{app:error_model}

In Sec.~\ref{sec:perf} we construct the residual error budget by separately measuring the induced bit-flip probability $p_1$ and induced dephasing probability $p_2$ per check. We now provide the model used to connect these quantities to the residual gate infidelity reported via ILRB.

We model the action of a single erasure check on the dual-rail logical subspace as a Pauli error channel,
\begin{equation}
\rho \;\to\; \left(1 - \frac{p_1}{2} - \frac{p_\phi}{2}\right)\rho
+ \frac{p_1}{4}\left(X\rho X + Y\rho Y\right)
+ \frac{p_\phi}{2}\, Z\rho Z,
\label{eq:error_channel}
\end{equation}
where $p_1$ is the induced bit-flip probability per check and $p_\phi$ is the induced pure dephasing probability per check.

After $N$ independent applications of the channel, the $Z$ expectation value evolves as $\langle Z\rangle_N = (1-p_1)^N$, so $p_1$ is simply the exponential decay rate of the longitudinal polarization per check. Experimentally, this corresponds to the decay of the difference $P(\1|\text{prep}\,\1) - P(\1|\text{prep}\,\0)$ with increasing number of checks at fixed evolution time (see Fig.~\ref{fig:fig3}(d)).

The total dephasing rate per check is $p_2 = p_1/2 + p_\phi$, where $p_1/2$ is the dephasing contribution from bit-flips and $p_\phi$ is the pure dephasing. When initializing in $+X$ state, the transverse coherence evolves as
\begin{equation}
\langle X\rangle_N = (1-p_2)^N = \left(1 - \frac{p_1}{2} - p_\phi\right)^N,
\end{equation}
making $p_2$ the exponential decay rate of phase coherence per check, measured via a spin-echo sequence with a variable number of checks at fixed evolution time (see Fig.~\ref{fig:fig3}(e)). The pure dephasing per check is then $p_\phi = p_2 - p_1/2$.

The average gate fidelity (defined with respect to the identity operation) is given by
\begin{equation}
F_{\rm avg} = \frac{1}{d+1}\left(1 + \frac{1}{d}\sum_k |\mathrm{Tr}(K_k)|^2\right)
= 1 - \frac{p_1}{3} - \frac{p_\phi}{3},
\label{eq:avg_fidelity}
\end{equation}
where $d=2$ for a single qubit, and the Kraus operators are $\{K_k\} = \{\sqrt{1-p_1/2-p_\phi/2}\,I,\;\sqrt{p_1/4}\,X,\;\sqrt{p_1/4}\,Y,\;\sqrt{p_\phi/2}\,Z\}$, in accordance with the channel defined in Eq.~\eqref{eq:error_channel}. The induced residual error per check is therefore $\varepsilon_{\rm ind} = 1 - F_{\rm avg} = p_1/3 + p_\phi/3$.

\section{Fluctuation of Erasure Lifetime Under Readout Drive}
\label{sec:terasure_fluctuation}
\begin{figure}[!]
\includegraphics[width = \columnwidth]{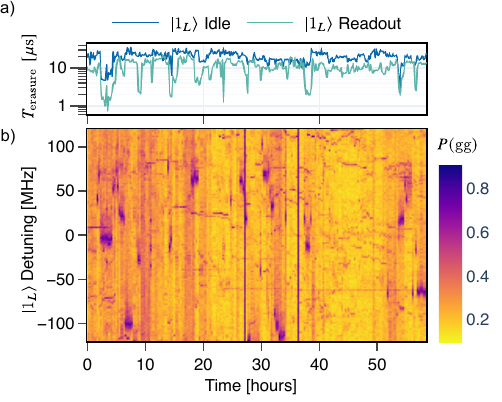}
\caption{\label{fig:fig_a2}\textbf{Correlation of fluctuating reduction in $\tera$ with near-resonant TLS.}
(a) Continuous monitoring of $|1_L\rangle$ $\tera$ with and without the readout probe.
(b) Interleaved flux-pulse spectroscopy mapping out fluctuating loss channels near $\g\leftrightarrow\1$ transition.
Frequencies with high probability of erasure event to $\g$ signify the presence of a loss channel which may be a TLS especially when the feature fluctuates temporally under an otherwise unchanging experimental configuration.
We note the sporadic reductions in $\tera$ frequently correlate with the appearance of near-resonant loss channels and the impact on $\tera$ during readout is more significant possibly due to readout induced sensitivity to off-resonant losses described in \cite{thorbeck2024readout, ziwen2026dispersive}. 
}\end{figure}
To correlate measurement induced erasure errors with qubit loss spectrum near the dual-rail hybrid mode transitions ($\g \leftrightarrow \0$ and $\g \leftrightarrow \1$), we map out the near-resonant loss channels with frequency swept fixed-time $\tera$ experiment \cite{lisenfeld2015observation}. The dual-rail qubit is initialized in one of the logical states then both transmons are flux-tuned simultaneously while in resonance to detune the hybrid mode frequencies. The dual-rail qubit is held at a detuned frequency for 1 $\mu$s before tuning back to the calibrated operating point for logical readout to determine survival of the excitation. We find that the fluctuation of $\tera$ is frequently correlated with near-resonant loss channels identified by interleaved flux-pulse measurements (see Fig.~\ref{fig:fig_a2}).

\section{Leakage Detection with SRO}
\label{app:leak}
\begin{figure}[!]
\includegraphics[width = \columnwidth]{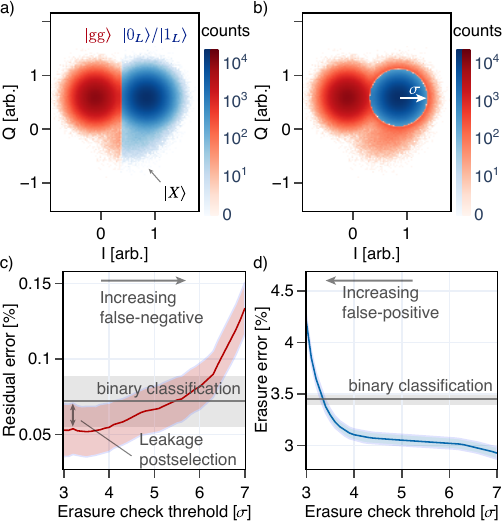}
\caption{\label{fig:fig_a3}\textbf{Detection and erasure conversion of additional leakage error with mid-circuit checks.} (a) Mid-circuit check readout signal distributions for an ILRB experiment where the calibrated erasure check is one where the readout is driven too strongly such that there is measurable but small probability of leakage to the higher excitation manifold of the dual-rail qubit as observed from the distribution of the readout signal beyond the two expected Gaussian distributions corresponding to the 0 and 1 excitation manifolds. The binary classification of the readout signal results in the majority of such leakage events being misclassified as being in logical subspace -- contributing to elevated and inaccurate residual error. (b) Alternative classification strategy by a circular threshold of $\sigma$ around distribution mean of the logical state readout allows us to better detect leakage events with mid-circuit checks (c,d) A sweep over the circular classification threshold shows the erasure conversion of residual error per check (c) from the leakage with more stringent threshold which comes at a cost of increasing erasure error per check in (d). }
\end{figure}

Since the SRO implements a direct dispersive readout that is sensitive to the excitation number of the dual-rail system, the same erasure check that distinguishes the logical subspace from $\g$ naturally provides visibility to leakage events involving higher excited states of the transmon pair, analogous to dispersive readout of higher transmon levels~\cite{wang2025probing}. States outside both the logical and erasure subspaces, such as the higher-excitation manifold populated by MIST events~\cite{sank2016measurement,khezri2023measurement,dumas2024measurement,dai2026characterization}, produce readout signals displaced from the expected logical and $\g$ distributions. This capability can be leveraged to convert leakage errors into detectable erasure errors using a modified classification strategy. When coherent control of the leaked state is available, detected leakage can additionally be actively reset mid-circuit; even absent such control, the classification alone serves as a useful gadget for leakage reduction via postselection.

To demonstrate this capability, we consider performing ILRB when the erasure check readout is intentionally driven above the optimal power such that there is a small but measurable probability of MIST to the higher excitation manifolds during each check. Under the standard binary threshold used for erasure classification, the readout signals from these leakage states (denoted $|X\rangle$) overlap predominantly with the logical-state distribution and are misclassified as remaining in the logical subspace (see Fig.~\ref{fig:fig_a3}(a)). This misclassification manifests as an elevated residual error in the ILRB characterization, as the leaked population is not flagged and instead contributes undetected errors to the postselected survival probability.

To better discriminate leakage events, one can design the readout probes and receiver to be more sensitive to the further dispersively shifted readout signal with techniques such as using a frequency comb of readout probes.
Here without modifying the readout probe, we simply adopt an alternative classification strategy for leakage detection in which the erasure check outcome is accepted as logical state only if the readout signal falls within a circular region of radius $\sigma$ centered on the mean of the logical-state distribution (see Fig.~\ref{fig:fig_a3}(b)). Readout signals outside this region (including those from both $\g$ and leaked higher-excitation states) are flagged as erasure events. By sweeping the classification threshold $\sigma$, we observe a clear trade-off between residual error and erasure error per check (see Fig.~\ref{fig:fig_a3}(c, d)): reducing $\sigma$ (tighter threshold) progressively converts leakage-induced residual errors into erasure errors at the cost of an increased false-positive rate and reduced postselection success probability.

This analysis highlights an advantage of the direct dispersive readout via the SRO: the same mid-circuit measurement that detects amplitude-damping erasure events with minimal back-action on the logical subspace can simultaneously serve as a leakage detection and erasure-conversion mechanism for higher excitation states. This dual functionality is available without additional circuit complexity, providing a path toward more complete error detection in erasure-based quantum error correction schemes where leakage errors would otherwise accumulate undetected and potentially lead to harmful spread of correlated errors~\cite{miao2023overcoming}.

\begin{figure}[!]
\includegraphics[width = \columnwidth]{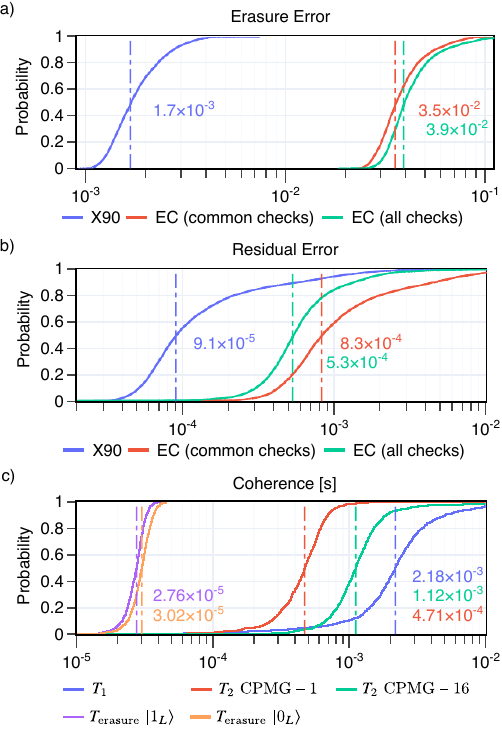}
\caption{\label{fig:fig_a4}\textbf{Stability of error rates and coherence.} Erasure error and residual error of X90 and erasure checks are monitored over 18 days with the cumulative distribution of the error shown in (a) and (b) respectively. Medians of the distributions are shown with dashed lines in the figures. Coherence of the dual-rail qubit is also monitored over a shorter duration of 5 days with the cumulative distributions shown in (c). No recalibration is done during the monitoring period. Sporadically, the performance and coherence would generally degrade as the TLS environment reconfigures and a TLS comes near resonance with the qubit. While this leads to a degradation, we note that the intrinsic flux noise suppression of the transmon dual-rail encoding along with the $\chi$-matching bandwidth provides the opportunity to recalibrate to a different operating point to avoid TLSes. 
}\end{figure}

\section{Stability of Erasure Check Performance}
\label{app:stab}
Dual-rail performance metrics fluctuate on a slow timescale much like the telegraphic noise observed for dual-rail frequency due to switching TLS defects in \cite{levine2024demonstrating}. 
The ILRB and induced error data presented in Section~\ref{sec:perf} were averaged over 1 hour and 18 hours, respectively. 
We additionally monitor the stability of coherence and infidelities over an extended period of time (5 days and 18 days, respectively) all without recalibration of the dual-rail qubit. See Fig.~\ref{fig:fig_a4} for the cumulative distribution of the coherence and infidelities.
We note that TLS defect reconfiguration is one of the leading causes of fluctuating performance which can be readily mitigated with the tunability of the dual-rail operating point made possible by the intrinsic robustness against single transmon flux noise sensitivity from the dual-rail coupling \cite{levine2024demonstrating} as well as the broad frequency of $\chi$-matching demonstrated in Section~\ref{sec:chi_dr}.

\FloatBarrier
\bibliography{references.bib}

\end{document}